\documentstyle[12pt]{article}
\begin{document}
\rightline{{\bf YCTP-P8-92}}
\rightline{Revised March 1992}
\baselineskip=18pt
\vskip 0.1in
\begin{center}
{\bf\large SOLAR NEUTRINO DATA AND ITS IMPLICATIONS}
\end{center}
\vskip 0.1in
\begin{center}
Evalyn Gates\footnote{\noindent Supported in part by the DOE.
Address after October 1: Department of Physics, University
of Chicago.},
Lawrence Krauss\footnote{\noindent Also Department of Astronomy.
Research supported in part by the DOE, the NSF and the TNRLC. Bitnet:
KRAUSS@YALEHEP}
and Martin White\footnote{\noindent Research supported in part by a
grant from the TNRLC.  Address after September 1: Center for Particle
Astrophysics, University of California, Berkeley.}

{\it Center for Theoretical Physics, Sloane Laboratory}

{\it Yale University, New Haven, CT 06511, USA}
\end{center}

\vskip .5in
\centerline{ {\bf Abstract} }

The complete and concurrent Homestake and Kamiokande solar neutrino data
sets (including backgrounds), when compared to detailed model predictions,
provide no unambiguous indication of the solution to the solar neutrino
problem.  All neutrino-based solutions, including time-varying models,
provide reasonable fits to  both the 3 year concurrent data and the full
20 year data set.  A simple constant B neutrino flux reduction is ruled out
at greater than the 4$\sigma$ level for both data sets.
While such a flux reduction provides a marginal fit to the unweighted averages
of the concurrent data, it does not provide a good fit to the average of the
full 20 year sample.  Gallium experiments may not be able to distinguish
between the currently allowed neutrino-based possibilities.

\newpage

Perhaps at no time in the past 20 years has there been more interest in the
solar neutrino  problem than at the present moment. The apparent deficit of
high energy solar neutrinos  observed by the Homestake Solar Neutrino Detector
over two decades \cite{Bahcall1} has now been confirmed by the Kamiokande
large underground water Cerenkov detector \cite{Hirata}.  Gallium
detectors are beginning to come online and the SAGE group has recently
published their first results \cite{SAGE} which seem to indicate a neutrino
deficit which cannot be explained by solar physics (for a brief discussion see
\cite{Krauss-SAGE}).
Other detectors are approved or are in the planning stages, and there is hope
that a solution to the solar neutrino problem may be at hand.
At the same time a growing number of theoretical neutrino-based ``solutions"
have been proposed.
Heading the list appears to be the MSW solution \cite{Mikheyev,Bahcall2},
which, in a restricted but reasonable range of neutrino masses and mixing
angles, allows significant reduction in the neutrino signal to be observed.
An alternative solution involves a large neutrino magnetic moment, either
diagonal or transitional, which causes neutrinos to oscillate into ``sterile"
partners while traversing the magnetic field of the sun \cite{Voloshin,Lim1}.
While this seems less theoretically compelling, especially in view of the
large neutrino magnetic moments required, it has the distinct advantage of
allowing not only the neutrino signal to be time varying over the solar cycle,
but also allows for a different time variation to be observed in different
detectors! \cite{Lim2}.

It may seem a priori that the simplest solution of the original Cl solar
neutrino problem resides in the solar model itself, namely that fewer high
energy neutrinos are created in the sun than the standard solar model
suggests.  It is important to determine if this possibility can be ruled out,
although it seems increasingly difficult to accommodate, especially in light of
the new SAGE results \cite{SAGE}.
Also, astrophysical mechanisms which reduce the high energy neutrino flux are
now not supported by any other solar measurements (most importantly the p-mode
fine structure \cite{pmode,Bahcall1}).
Meanwhile, several studies incorporating recent Cl data into the 20 year
observations provide very tempting, if not compelling, evidence of time
variations in the Cl signal \cite{Bahcall3,Krauss2,Bieber} which may be
correlated, in some yet to be determined way, with the solar cycle.
On the other hand, the Kamiokande data appears naively to show no such time
variation.
The presence, for the first time, of two different data sets for the solar
neutrino signal should allow a number of finer tests of solar neutrino models
to be made (see e.g. \cite{Hirata}).
Surprisingly, however, rarely have all the data been used.  For example,
analyses have been performed comparing the Homestake 20 year
``average" signal with the averaged Kamiokande 3 year signal.   It is not
clear that such a procedure is correct.  Until we have a better idea of
what is at the root of the solar neutrino problem, we can make no
{\em a priori} claims about what the Kamiokande signal would have been if the
detector had also taken data during the 20 years Homestake was operational,
especially given the apparent variations in the Cl data during this period.
All of the data points from both experiments should be exploited, and the
error bars examined.  For guidance on how to treat the entire data sets one
can first analyse the data during the period in which the two detectors were
both running concurrently.  It is only during this time that we have a direct
independent check on the Cl data, and can check for consistency between the
data sets. One might then be guided on how to use all the data to test
various hypotheses.  This is the spirit of the following work.  We have
utilized the entire Homestake and Kamiokande data sets, concentrating first
on the concurrent data sets and then on all the data, in order to investigate
the range of models which may or may not fit the data.  We have carried out
extensive numerical model calculations, in which neutrinos are propagated,
with complete phase information, through much of the sun, in order to
estimate the flux in various neutrino species at the earth's surface.
We have also used realistic models of detector sensitivity in order to turn
fluxes into detection rates.

We emphasize that using the concurrent Kamiokande data to ``check" the Cl data
is important beyond the strict question of whether or not any solar cycle
time variation exists.  It allows us to understand how best to treat the full
20 year Cl data to explore solutions to the solar neutrino problem. We
display in figure 1 (a) and (b) the two full data sets, and
the concurrent data sets. Both data sets are normalized to the Standard Solar
model \cite{Bahcall1} predictions (see section (3)).

While quoted averages of the two data sets appear at first sight to differ,
when the full data sets are displayed this issue is less clear.  The Cl signal
clearly has much more jitter, with several apparently anomalously low points,
but aside from this one might not be surprised if told that all data came from
a single detector.
This may suggest that a simple, energy independent deficit of B neutrinos
could be consistent with all of the data.  To properly explore this
possibility, as well as the possibility that the solar neutrino deficit is
neutrino related, a more quantitative approach is required.

\vspace{.2in}
\noindent {2. \bf Neutrino flux at the earth}

The neutrino spectrum predicted by the Standard Solar Model (SSM) is
described in detail by Bahcall in \cite{Bahcall1}.  The dominant neutrino
flux, that due to the pp reaction in the sun, with energies less than
0.42 MeV, is unobservable in both the Cl and Kamiokande detectors, due
to their thresholds.  The component of the flux which gives the dominant
contribution to the Cl signal, and the entire contribution to Kamiokande
is the high energy $^8B$ continuous spectrum
($^8B\rightarrow ^7Be^{*}+e^{+}+\nu_e$),
with neutrino energies up to $15$MeV and a total predicted flux at the earth of
$(5.8\pm 2.2)\times 10^6 \mbox{cm}^{-2}s^{-1}$  (``$3\sigma$" theoretical
error).
The only other component of the neutrino spectrum contributing to the Cl
signal at greater than the $5\%$ level are the $Be$ neutrinos
($^7Be + e^{-}\rightarrow ^7Li + \nu_e$), with fixed energy 0.862 MeV and a
predicted flux of $4.7\pm .7(3\sigma)\times 10^9 \mbox{cm}^{-2}s^{-1}$.

There are two ways one might expect to alter these predicted fluxes.
First one might lower the overall flux by a fixed amount by postulating some
new solar physics.  For example, if the core temperature is lowered compared
to the SSM, the B signal can be significantly reduced (such a temperature
reduction is the aim of many non-standard solar models,
e.g. see \cite{Bahcall1}).

We have incorporated these possibilities in our analysis by treating
the net B flux as a free parameter in one set of runs, and examining the
goodness of fit with the combined data sets as this parameter is varied
compared to the SSM.  While this is a very simplistic ``non-standard solar
model" we can use it to perform straightforward statistical tests of
how well the data is fit by models aiming at such a B flux reduction.

The other possibility is that the origin of the solar neutrino problem
lies in the properties of neutrinos themselves.  If neutrinos have non-zero
mass eigenstates which do not coincide with weak eigenstates, neutrino
propagation will lead to oscillations between the different weak states,
namely between electron, muon, and tau neutrinos.  Since the Cl detector
is sensitive {\em only} to electron neutrinos, while the Kamiokande water
detector is sensitive {\em predominantly} to electron neutrinos, such
oscillations could have the possibility of reducing the observed signal in
both detectors.
Moreover, the presence of matter can enhance the oscillations between neutrino
species \cite{Mikheyev} due to the presence of level crossings which occur as
the background electron density varies.  If one supplements neutrino masses
with large magnetic moments, which in general need not be diagonal in the weak
basis, then another possibility arises.  Magnetic fields in the sun could
cause oscillations between left and right handed neutrino states, with or
without induced level crossings \cite{Voloshin}. In general, left-right
mixing can allow neutrino states to oscillate into antineutrino states,
unlike the pure MSW mechanism \cite{Lim1}.  In any case, as long as the right
handed states have suppressed interaction rates in the detectors, this can
reduce the observed neutrino signal.  Moreover, it allows for a possible
correlation with the solar cycle, although the required neutrino magnetic
moments, at least for currently envisaged magnetic field strengths in the
sun, are large enough to cause other potential astrophysical problems
\cite{Raffelt}. Finally, in the most general case, both effects may be
operational with the different factors dominating in different regimes of
mass, mixing angle, and magnetic field space \cite{Lim1}. This allows
independent time variations to be observed in the two detectors \cite{Lim2},
and it is this general case which we shall consider here.

We followed explicitly the propagation of neutrinos through the sun by
numerically integrating the Hamiltonian evolution equation for
neutrinos through matter for a two generation model with Majorana-type
transition magnetic moment and off diagonal mass terms
\cite{Lim1,Lim2,Press,Rosen}.
\begin{equation}
i {d\over dt} \left[
\begin{array}{c}
  \nu_e \\ \nu_{\mu} \\ \bar{\nu}_e \\ \bar{\nu}_{\mu}
\end{array}
\right] = H \left[
\begin{array}{c}
  \nu_e \\ \nu_{\mu} \\ \bar{\nu}_e \\ \bar{\nu}_{\mu}
\end{array}
\right]
\label{eqn:hamilton-equation}
\end{equation}

The Hamiltonian for the system \cite{Lim1} is given by
\begin{equation}
H=\left[
\begin{array}{cccc}
a_{e} & {\Delta m^2\over 4E_{\nu}}\sin 2\theta & 0 & \mu B \\
& & & \\
{\Delta m^2\over 4E_{\nu}}\sin 2\theta &
  {\Delta m^2\over 2E_{\nu}}\cos 2\theta + a_{\mu} & -\mu B & 0 \\
& & & \\
0 & -\mu B & -a_{e} & {\Delta m^2\over 4E_{\nu}}\sin 2\theta \\
& & & \\
\mu B & 0 & {\Delta m^2\over 4E_{\nu}}\sin 2\theta &
  {\Delta m^2\over 2E_{\nu}}\cos 2\theta - a_{\mu} \\
\end{array}
\right]
\end{equation}
where B is the magnetic field,
$a_{e}=G_F(2N_{e}-N_{n})/\sqrt{2}$ and $a_{\mu}=G_F(-N_{n})/\sqrt{2}$\
with $N_{e},N_{n}$ the electron and neutron densities as a function of
radius in the solar interior.
We used the following fit to the electron and neutron densities in the
standard model sun \cite{Bahcall1}
\begin{equation}
N_e = \left\{ \begin{array}{cccccc}
 2.45\times 10^{26} \exp\left( -10.54 x \right) &
 \ 0.2 & < & x & < & 1 \\
 6\times 10^{25} \left[ 1 - 10x/3 \right] /\mbox{cm}^3 &
 \ 0.1 & < & x & < & 0.2 \\ \end{array} \right.
\end{equation}
\begin{equation}
N_n = \left\{ \begin{array}{cccccc}
 2.45\times 10^{26} \exp\left( -10.54 x \right)  &
 \ 0.2 & < & x & < & 1 \\
 2\times 10^{25} \left[ 1 - 21x/5 \right] /\mbox{cm}^3 &
 \ 0.1 & < & x & < & 0.2 \\  \end{array} \right.
\end{equation}
where $x=r/R_{\odot}$.

The free parameters in the calculation are the neutrino energy E, mass-squared
difference $\Delta m^2$, vacuum mixing angle  $\sin^2(2\theta)$ and
Zeeman energy $\mu B$: the product of the transition magnetic moment and solar
magnetic field.  For reasons of simplicity we assumed this to be uniform over
the radiation and convection zones in the sun, falling sharply to zero
at the exterior.

The evolution in the interior was performed using a Runge-Kutta
algorithm with adaptive step size control \cite{Press} in double precision
arithmetic.  On order of $10^5$ steps were taken for the higher mass gaps, the
neutrinos being evolved from just before the resonance \cite{Rosen}
\begin{equation}
  \left. \sqrt{2} G_F N_e \right|_{res} = {\Delta m^2\over
2E}\cos(2\theta)
\end{equation}
to the edge of the Sun.

In the exterior of the sun, where the magnetic field is assumed to be zero,
the neutrino and anti-neutrino sectors decouple and the vacuum oscillations
can be computed using standard analytic formulae (see \cite{Rosen}).
Since each detector signal averages over a period of at least 2 months (though
not necessarily weighting times evenly)  we modelled  the motion of the Earth
in a simple way by averaging the vacuum oscillations over an Earth-Sun
distance of $d(1-e/2)$ to $d(1+e/2)$, where the semi-major axis is
$d = 1.496\times 10^8$km and the eccentricity of the Earth's orbit is
$e=0.0167$.  This corresponds to the variation in the Earth-Sun distance
over 3 months.

The general form of the propagation matrix for neutrinos allows for the
conversion of electron neutrinos into muon neutrinos and also into muon
and  electron  antineutrinos.  The latter conversion can occur in two steps,
either by a magnetic moment induced oscillation followed by an MSW type
oscillation, or the reverse. Assuming initially electron neutrinos are
emitted, the probability $P_i$ of finding each of the 4 species at the
Earth was computed for a grid of the 4 parameters.  For the continuum spectra
we calculated the probabilities for 30 energies ranging from $0.5MeV$ to
$15MeV$ in $0.5MeV$ steps, and we also calculated the probabilities at
$0.862MeV$ and $1.442MeV$ corresponding to the $^7$Be and pep neutrino lines
respectively. The mass gap, $\Delta m^2$, ranged from $10^{-5} eV^2$ to
$10^{-8} eV^2$; for  higher mass gaps the Zeeman energy plays no role and
pure MSW/vacuum  mixing, results.  This case has been well studied and the
higher mass gaps in the so called ``adiabatic regime" may already be ruled
out by  experiment \cite{Hirata}.
The vacuum mixing angle, $\sin^2(2\theta)$, ranged from $0.01$ to $1.00$
and the Zeeman energies, $\mu B$, from 0 to $5\times 10^{-10} \mu B$kG.
The best limit on neutrino transition moments is astrophysical, coming
from the luminosity of red giant stars before and after the He flash
\cite{Raffelt},
\begin{equation}
 \mu < 3\times 10^{-12} \mu_B\ (3\sigma)
\end{equation}
and the best lab limits (from $\bar{\nu}_e-e$ scattering) are
\cite{Marciano-review}
\begin{equation}
|\kappa_e| < 4\times 10^{-10},\ |\kappa_{\mu}| < 10^{-9};
\ \  \mu_i = \kappa_i \mu_B
\end{equation}
so the larger Zeeman energies require enormous fields in the solar interior.

The expected event rates in the detectors were calculated by convolving
known neutrino cross sections with published detector efficiencies
\cite{Bahcall1,Krauss2,Nakahata}.

\vspace{.2in}
\noindent {3. \bf The Data}

The 90 Homestake data points between the years 1970 and 1991 were obtained with
about 2 months of integration time per point.  The time shown for each
Homestake data point in figure 1 is the mean time of production
of the radioactive Argon atoms (see \cite{Bahcall1} for a description).
For each point the experiment reported an upper limit on the production
rate, a lower limit on the rate, and the mean value of the rate, all
determined by a maximum likelihood fit to the data \cite{Cleveland}. The
errors about the mean were generally symmetric, except in the case where the
lower limit would have become negative, in which case the reported error bars
were quoted as half the difference between the upper limit and zero, and were
thus sometimes artificially small.
This suppression of the errors would artificially increase the weighting
of these points in any fit to the data.  In order to remove this effect, we
utilized fully symmetric error bars on all points.   The size of $1\sigma$
error bars was fixed to be the difference between the reported upper limit and
the mean value for each point. It has been {\em calculated} that $.08\pm .03$
argon atoms/day are produced \cite{Bahcall1} by the (muon induced) background.
In determining the average Homestake signal it is appropriate to subtract
this background {\em after} the average Ar rate has been computed from the
total data set, and add errors in quadrature.  When performing
a point by point fit of theory to the data, however, it is appropriate to
subtract this background from each data point and add its uncertainty to the
rate uncertainty for each point in quadrature.\footnote{The average rate
(which converts to $.26\pm .04$ SSM) quoted by the Homestake group comes from
a maximum likelihood fit of N=61 runs to a constant background plus 1 decaying
species.  The division of the counts into (counter) background and signal is
different if the runs are analysed separately or collectively, the (counter)
background in a run by run analysis being quite variable. We calculate our
average rate as the average of the values quoted per run, for N=90 runs. Note
that $.26$ is bracketed by our weighted and unweighted values.  The larger
error, .04, is consistent with the smaller number of runs analysed by the
Homestake group.}
Figure 1 displays the values divided by the standard solar model (SSM)
predicted rate.   Because of the unusually small errors on many of
the points with small rates, the treatment of errors in the Homestake
experiment has been an issue of some debate.  In particular the ``error"
determined by the maximum likelihood fit is not a Gaussian $1\sigma$ error for
points with small numbers of counts $(N\le 5)$ and the use of a $\chi^2$
analysis will not weight these points correctly (see e.g. \cite{Filippone}).
To consider the effect of this, for the analysis of the non-standard
solar models and the MSW neutrino model we also used the method of
\cite{Filippone} to analyze the Homestake data, while still using
$\chi^2$ for the Kamiokande data.

The Kamiokande data is more straightforward.
Over the period 1987-1990, five data points have been obtained, based on real
time measurements of the directional solar neutrino signal, averaged over a
period of several months.  These data points, along with errors, were
presented as a fraction of the rate predicted by the SSM \cite{Hirata}, and,
as shown in figure 1, were used directly in this analysis.

Finally, we decided not to additionally weight the Kamiokande and Cl data
points in terms of the length of the measuring interval associated with each
point.  In the first place, longer runs in the Cl experiment do not mean more
data.  Because the produced Ar atoms decay with a 35 day half life they will
eventually reach an equilibrium abundance after several months exposure.
Secondly, the small error bars on the Kamiokande data points presumably
reflect the longer exposure times for each point in this experiment, and thus
measuring time will in this case naturally be taken into account in any
weighting by errors of the data.

\vspace{.2in}
\noindent {4. \bf Analysis}

In an effort to determine how the current solar neutrino data constrains the
various possible models discussed in (2) we compared the
predicted signals in both detectors to the data by means of a $\chi^2$
goodness-of-fit procedure.
For each model we computed the predicted signal over a range of model
parameters, and for each combination calculated the value of $\chi^2$ for the
signal compared to the data.  We then examined the parameter space for $\chi^2$
values corresponding to confidence levels of $68\%$ and $95\%$.

The different models we considered are:
\begin{itemize}
\item[(a)] Non-standard solar model: B flux reduction
\item[(b)] Non-standard solar model: (B+Be) flux reduction
\item[(c)] Neutrino Mass model (no magnetic moments-constant flux)
\item[(d)] Neutrino Mass model (with magnetic moments-variable flux)
\end{itemize}

In case (c), each combination of the neutrino mass-squared difference and
vacuum mixing angle produced a constant fit to each of the
detector signals.  In case (d) in addition to these parameters, the quantity
$\mu B$ was assumed to have the form $\mu B = A  + C f(t)$.  In this case
$f(t)$ was set to either $\cos(\phi + kt)$, where $\phi$ and $k$ were
determined from sunspot data, or to a sawtooth function of unit amplitude
with a net period equal to the solar cycle and the position of the cusp given
by time $\tau$.
This latter model was chosen based on an earlier suggestion by Bahcall and
Press \cite{Bahcall3} that the neutrino time variation could be well described
by such a function. We considered $\tau=8.05$ years, based on their fit to the
Ar data, and $\tau=6.65$ years based on their fit to sunspot data.  Thus in
case  (d) there are two additional parameters, A and C, involved.
The translation of $\chi^2$ values into confidence levels depends upon number
of degrees of freedom.  In determining the goodness of fit of models (a,b,c)
with various sets of parameters, the number of degrees of freedom was set
equal to the number of data points, since the model predictions are fixed
once the parameters are fixed, and no parameter in this test is minimized to
fit the data.  In model (d) the number of degrees of freedom was reduced by 2
since A and C were fit to the data before goodness of fit was evaluated.

Our results are displayed in Tables \ref{tab:sn-neutrino},\ref{tab:sn-nssm}
and figures 2 - 7.
The tables list the ``best fit" (i.e. smallest $\chi^2$) model parameters
along with degrees of freedom (df)\footnote{Note the number of degrees of
freedom to be used for a goodness-of-fit and the number quoted for a ``best
fit" are not the same, the latter being smaller by the number of parameters
varied in the fit.} for fits to:
\begin{itemize}
\item[(i)] the concurrent 3 year data, and the averaged 3 year data,
\item[(ii)] the complete 20 year data set, and the averaged 20 year data
\end{itemize}
We do not place much significance on the actual value of the best fit
parameters, rather we would emphasize the regions in
$\Delta m^2-\sin^2(2\theta)$ space for which the model fits the data at
a given confidence level.

Let us review the fits to each of the data sets in turn.

\vspace{.1in}

\noindent (i) {\bf Concurrent data set:}
In spite the apparent similarity of the two signals during this period, the
simplest apparent resolution of the solar neutrino problem, that obtained by
reducing the B neutrino flux alone, is ruled out at greater than the $4\sigma$
level based on a comparison with the weighted data points (including a Be
reduction by the same amount allows a fit at the $3\sigma$ level).
This discrepancy is because the small error bars on the low Homestake points
heavily skew any fit.  The mean value of the Homestake data during this period
rises  from .25 to .36 of the SSM prediction if each point is equally
weighted and the fit to a non-standard solar model improves dramatically.
In this case, if the SSM B flux is reduced by a constant factor, the fit
to the unweighted averages is acceptable over a small range at the 99\%
confidence level. Whether or not to ignore the heavy weighting of the apparent
anomalously low Cl data points therefore becomes an important issue if one is
to claim non-standard solar models are ruled out by the combination of Cl and
Kamiokande data, at least during the period in which the data was taken
concurrently.
If the procedure of \cite{Filippone} is used the non-standard solar model
just fits the concurrent data at the 99\% confidence level\footnote{The
procedure of \cite{Filippone} makes use of the likelihood ratio test in which
the test statistic is $\chi^2$ distributed in the limit of a large number of
data points.  In applying this test to the concurrent Homestake data we should
bear in mind that there are only 20 data points.}, with the favored boron flux
reduction at 37\% of the SSM.

If no model fit the complete fully weighted concurrent data sets, this
would provide strong evidence in favor of the assumption that the jitter in
the Cl signal precludes its use directly in constraining models, and might
provide motivation for ignoring the quoted error bars on the data.  As can be
seen, however, all the models with neutrino masses, including those with a
time variability, provide reasonable fits to the data (at 95\% confidence
level).
The range of fit of the MSW model to this concurrent sample is shown in
figure 2 (a), along with the claimed fit to the 20 year
averaged data by Bahcall and Bethe \cite{Bahcall2} (solid line).  We see that
the Bahcall and Bethe line passes through the arm of the 95\% confidence level
region.
If the unweighted averages of the Homestake data sets and the Kamiokande
average rate are compared to the MSW prediction, the allowed regions are
shown in figure 2 (b).
Notice that the fit to the unweighted average is good at the 68\% level over a
range of parameters and the 68\% region coincides with the Bahcall and Bethe
best fit line.  An almost identical region is obtained for the fit to the
weighted averages of the data, suggesting the poorer fit in the case of the
individual points is due to ``jitter" in the data.
If the analysis is done using the method of \cite{Filippone} the MSW model
still fits, though the goodness-of-fit is slightly worse than for the case of
the straightforward $\chi^2$ fit.

We now switch to the time dependent fits, involving a non-zero transition
magnetic moment. The ``best fit" magnetic field peak Zeeman energy has a value
of $4.6\times 10^{-10}\mu B$kG for the cosine and $4-5\times 10^{-10}\mu B$kG
for the sawtooth fits, which are essentially as good as the MSW fits. Because
the 20 year data provides more compelling evidence of time variability, we
also investigated the goodness of fit of the 20 year ``best fit" parameters to
the 3 year concurrent set in the time varying models.  The ``best fit" values
differ somewhat from the best fit to the 3 year data, but they are still
comparably good. This indicates that there is no evidence from the concurrent
data against the same time variation inferred from the 20 year Cl sample.

\vspace{.1in}

\noindent (ii) {\bf 20 year data set:}
A non-standard solar model doesn't fit the full data much worse or much better
than the 3 year data. The disagreement with the complete weighted data sample,
allowing only the B flux to be reduced (in this case to 0.1 SSM!), is still at
$\approx4.5\sigma$.  Now however that now the disagreement with the unweighted
average rate (requiring a flux reduction to 0.15 SSM) is comparably bad.
Allowing the Be flux to change as well reduces the disagreement, but the fit
to the unweighted average in this case is at best only marginal (99\%
confidence level).
The procedure of \cite{Filippone} decreases the goodness-of-fit dramatically,
with the best fit (at 20\% of the SSM boron flux) ruled out at $>5\sigma$.

The MSW model fit to the 20 year data is shown in figures 3 and 4.
Notice the line of best fit is shifted slightly from the Bahcall and Bethe line
due to the inclusion of the latest Homestake data but the fit is still good at
the 95\% confidence level.
The fits to the weighted and unweighted averages are good (better than 68\%)
as one might expect.
If the method of \cite{Filippone} is used to compute the $\chi^2$, thus
taking account of the Poisson statistics of the low points, the best fit is
only acceptable at the $\approx 5\sigma$ level!
The fact that both the non-standard solar model and MSW fits, in which the
prediction is a constant, are worse using the method of \cite{Filippone}
than using a normal $\chi^2$ procedure suggests that this latter method is
much more sensitive to ``jitter" in the data.

Since it is perhaps the simplest and most elegant of the proposed neutrino
based ``solutions" to the solar neutrino problem we feel the MSW model
deserves a closer inspection.  In this regard we have developed a new way of
presenting the comparison between theory and observation. For the 680
$(\Delta m^2,\sin^2 2\theta)$ parameter pairs we calculated in our
study, we display in figure 5 a plot of the MSW predictions for Homestake vs
Kamiokande.  While {\em a priori} one might expect such a plot to ``fill"
much of the plane, one can see that the allowed region is in fact a narrow
band passing from bottom left to top right. This behaviour is due to the fact
that high energy $^8B$ electron neutrinos make up most of the signal for both
detectors, leading to a strong correlation in the signals for an energy
dependent $\nu_e$ flux reduction.  (We thus expect that adding the neglected
contributions from $^{15}O$ and hep neutrinos to the Homestake signal will
broaden this band slightly.)
Still the narrowness of the band is a surprising indication of the strong
constraints on the predictions of the MSW solution.  Also shown in figure 5
are the averages of the actual rates seen in the detectors.  In this way one
can obtain a clear and immediate graphical picture of how well the MSW
solution as a whole can reproduce the observed averages. As can be seen, the
fair overlap between (the constrained) theoretical phase space and the
observations is suggestive.

The low points in the pre-1987 sample can be well accommodated, as has been
previously noticed, by a time varying neutrino signal. In addition, as
stressed earlier, resonant spin-flavor transitions also allow ``arbitrary"
Kamiokande time variation for a given variability in the Cl data.  As expected,
therefore, we find that the complete data sample can be well fit over a wide
range of parameter space by a time varying magnetic field coupled with a large
neutrino transition magnetic moment.  Shown in figure 6 (a) and (b) are the
regions of mass-mixing angle space allowed at the $68$ and $95\%$ confidence
levels when the magnetic field time variation is fixed at the value which
provides the minimal $\chi^2$ fit to the data for a (a) cosine or (b) sawtooth
time dependence.  (The actual region of parameter space allowed in this case
is a 4 dimensional space in mass, mixing angle, and magnetic field time
variation -- difficult to  draw, but whose boundary in the extreme limit of
zero magnetic field splitting would reduce to the MSW plot already presented.)
The cosine fit to the data at this optimum magnetic field value is obviously
better than the zero field MSW fit, while the sawtooth fit is even broader,
and slightly better than the cosine fit at the optimum magnetic field value.

The apparent jitter and/or the occurrence of anomalously low data points in
the Cl data sample, which dominates over the Kamiokande sample in the 20 year
fits (by about 4 to 1 in the $\chi^2$ determinations), cannot be dismissed
based purely on statistical grounds alone.  We have investigated whether one
might be forced to ignore or rescale the error bars in order to reduce this
effect by examining the variance of both the weighted and unweighted Cl 20
year samples. The mean value of the of the Cl signal for the complete 20 year
weighted sample is $1.70\pm .22$ SNU.  This is significantly smaller than the
unweighted average of $2.21\pm .24$ SNU.  Nevertheless, the $\chi^2$ per degree
of freedom for this weighted average is 1.07.  This indicates that there is no
necessity to rescale errors to account for the variance of the sample from the
mean.  Alternatively, the unweighted sample has a mean variance per point of
1.7 SNU.  This is comparable to the error per point in the weighted
sample, indicating again that there is no evidence that the errors are
skewed in any way.

Finally we stress a somewhat non-intuitive result.  In the 20 year
sample, the Cl data clearly dominates in any fit.  One may feel that comparing
model predictions to average values may alleviate this problem by treating the
two data sets with equal weight.
However, the relative errors determined for the Homestake mean values are
small enough so that the Homestake result dominates the fit to average values
(weighted or unweighted) more than it does a fit to the complete sample.
Thus, if the Cl data is suspect, for any reason, using average values rather
than the full data set will only exacerbate this problem.

One way in which we might hope to proceed further in distinguishing between
models is to examine the predictions for the Ga solar neutrino experiments
(SAGE and GALLEX collaborations) which are currently beginning to run.
Estimates of gallium rates predicted by the models we have considered are
summarized in the last two columns of Table \ref{tab:sn-neutrino}.
For a given model, we have computed the range of neutrino rates that would
be seen in a Ga-based detector for the region of parameter space not already
excluded at the 68\% and 95\% confidence levels by the present Homestake and
Kamiokande data.  The time dependence of the predicted Gallium rates for the
time-dependent models varied widely (including no significant time variation)
for equally allowed parameter sets. Thus measuring the time dependence of the
rates in Gallium detectors might help further constrain these models, although
if uncertainties in the data are on the same order as the Cl data, a clear
measurement of time dependence is unlikely in the short term.  Moreover, an
observation of no time variation in the Gallium detectors would once again
not provide definitive evidence against time variation in the Cl signal.
In the context of neutrino based models then the SAGE result,
$(20\pm 38)$ SNU, is perhaps the least enlightening result one could obtain
from a theoretical point of view.

Kamiokande itself now provides another constraint on resonant spin-flavor
conversion models.
Electron neutrinos can be converted to electron anti-neutrinos in the sun,
and these contribute to the isotropic background signal in the Kamiokande
detector.  Thus, the flat background of isotropic events seen by the
Kamiokande detector can place a limit on the flux of electron anti-neutrinos
\cite{Lim2}. Although a careful analysis of the data in this regard
has not yet been performed, estimates of the flux of electron anti-neutrinos
for neutrino energies greater than or equal to 10.6 MeV for the time period
June 1988 through April 1989 are less than approximately 10\% of the expected
electron neutrino flux predicted by the SSM \cite{Barbieri}. For the models
discussed in this paper, the predicted electron anti-neutrino fluxes ranged
from 0 to 30\% of the SSM $\nu_e$ flux.  Figure 7
outlines regions of parameter space excluded for various flux limits, for
Zeeman
energies of $2.0\times 10^{-10}$ and $5.0\times 10^{-10} \mu B$kG respectively.
(Indicative of average and peak Zeeman energy values which appear in the best
fit solutions.)  Note that some regions favored by the time varying models are
eliminated by the 10\% -of-background cut, but none of the time varying models
are completely eliminated on the basis of this constraint alone.  As the
energy threshold for the Kamiokande background subtraction is reduced, more of
the parameter space for magnetic moment induced oscillations can be probed.
However, it is worth noting that our results suggest that none of the present
``allowed regions" for the time varying models would be eliminated even
if a background cut at the 5\% level were made.  It is possible that the SNO
heavy water detector may eventually be able to distinguish the antineutrino
signal more clearly from the neutrino signal, and thus could further improve
these bounds.

\vspace{.2in}
\noindent {5. \bf Results}

For convenience we summarize the above analysis and restate the main results:
\begin{enumerate}
\item Non-standard solar models which result in a reduced boron flux
are ruled out, for the concurrent weighted data sample, at the $4\sigma$
confidence level.  This limit is basically unchanged when the rest of the Cl
data is taken into account, though the required flux reduction is more extreme.
If the unweighted Cl average signal is utilized instead, this simplest
non-standard solar model fits at the 98\% confidence level for the concurrent
data sample.
In this case, however, the fit to the unweighted average of the full 20 year
sample is {\em incompatible} at the $\approx 4-5\sigma$ level, due to the low
long-term Homestake average.
The SAGE results now also appear to argue against this possibility.
\item The MSW neutrino mass solution of the solar neutrino model over much of
the range claimed by Bahcall and Bethe fits the concurrent and 20yr weighted
data at only the 95\% confidence level.  We have no statistical evidence that
the error bars in the Cl data are anomalous, but if the unweighted mean is
utilized instead, the MSW fits improve significantly. This suggests the jitter
in the Homestake data may be the cause of the higher $\chi^2/$dof.
On a Homestake vs. Kamiokande plot the MSW prediction appears as a thin band
which overlaps the averaged data.  In this way, the agreement between theory
and averaged data is more easily pictured.
\item Models with resonant spin-flavor
conversion due to a varying magnetic field in the sun fit the data with a
confidence level which is at least comparable to the MSW fits -- even for the 3
year concurrent sample in which no time variation in the Kamiokande signal is
obvious.  As expected, the  time-varying models provide acceptable fits to the
complete weighted data set  much more broadly than the MSW models do, and in
the
case of a sawtooth  time-dependence the best fit is also greatly improved. The
maximum Zeeman splitting needed in these cases is rather large, of order
$2-5\times 10^{-10}\mu B$kG. \item Most neutrino based solutions to the solar
neutrino problem not excluded at the 95\% confidence level predict roughly
comparable rates in Ga, between 5-65 SNU.  Non-standard solar models which are
not excluded predict rates greater than 90 SNU.  Hence, Ga can decisively rule
out non-standard solar models, but cannot distinguish well between neutrino
based solutions.   Acceptable time-varying models predict a wide range of
possible time variation in Gallium, including almost no observable variation.
\item Kamiokande can restrict the allowed parameter range for spin-flavor
conversion models, and already rules out $\Delta m^2$ in the range
$10^{-8}-10^{-7} eV^2$, for mixing angles greater than
$\sin^2(2\theta)\sim 0.3$.  This limit comes from the isotropic background in
the experiment and will improve with time. The SNO detector might improve
these further.
\end{enumerate}

\vspace{.2in}
\noindent{6. \bf Conclusions}

The Kamiokande experiment can provide a useful check on the Homestake
experiment, and the combined data from both experiments during their
concurrent running is consistent with a wide variety of models.
Unfortunately, however, the specifics of which model and what parameters
appear to be favored depend upon how one treats the data, so that no
categorical conclusions can yet be made.

Future experiments at Kamiokande and with Ga may not allow much finer
distinctions between neutrino-based models to be made, but they could
definitively rule out non-standard solar model based solutions of the solar
neutrino problem. At this point 20 years of experiments have at least firmly
established the existence of the solar neutrino problem and pointed to new
microphysics as the likely solution. To gain the information necessary to
completely  resolve this issue it will be necessary to measure the solar
neutrino spectrum  itself.  If neutrino mixing is indeed the cause of the solar
neutrino problem then a knowledge of which energies are most suppressed would
give us a better handle on the underlying mechanism and parameters (for example
in simple MSW mixing, in the regions considered here, lowering $\Delta m^2$ for
a given mixing angle lowers the threshold energy below which
$\nu_e\rightarrow\nu_x$ conversion takes place).

Experiments with this goal in mind (i.e.\cite{cabrera,wilczek}) are important
to
pursue.  In this way a new window on physics at scales beyond those accessible
at present accelerators may be fully explored.

We thank Ken Lande for providing us with the complete sets of Chlorine
neutrino data and for useful discussions on both the Cl and Ga experiments, and
M. Smith for informing us of the work of Filippone. We also thank C. Baltay for
useful discussions, P. Langacker for helpful advice on error handling, and D.
Gelernter and D. Kaminsky of the Linda group of the Department of Computer
Science at Yale for running our evolution code on their complex.

\clearpage

\begin{table}
\begin{center}
\begin{tabular}{llc}
& & \\
Experiment  & Averaging Method          & Average \\ \hline
Kamiokande: &                           & $0.4600\pm 0.0781$ \\
Homestake:  &                           &                    \\
            & 20 year weighted:         & $0.2153\pm 0.0284$ \\
            & 20 year unweighted:       & $0.2799\pm 0.0309$ \\
            & concurrent weighted:      & $0.2475\pm 0.0436$ \\
            & concurrent unweighted:    & $0.3602\pm 0.0528$ \\
\end{tabular}
\end{center}
\caption{Average values for solar neutrino data}
\label{tab:sn-averages}
\end{table}

\clearpage

\begin{table}
\begin{center}
\begin{tabular}{lcccc}
& & & & \\
Model & $\chi^2$ (d.f.) & Parameters$^{*}$ & Ga(68\%) & Ga (95\%)\\
\hline
\multicolumn{3}{l}{\bf Concurrent Data:} & &  \\
MSW             & 32.6(23) & 1.58,0.25,---,---  &       & 5-56 \\
Cosine          & 30.9(21) & 1.26,0.10,2.3,2.3  &       & 5-66 \\
Sawtooth (6.65) & 31.7(21) & 0.25,0.20,2.0,2.0  &       & 5-66 \\
Sawtooth (8.05) & 31.0(21) & 0.16,0.45,2.4,2.4  &       & 5-66 \\
Cos (20yr)      & 31.7(23) & 1.58,0.20          &       &      \\
Saw (20yr-6.65) & 32.1(23) & 1.58,0.15          &       &      \\
Saw (20yr-8.05) & 31.4(23) & 1.26,0.10          &       &      \\
\multicolumn{3}{l}{\bf Concurrent Data (averages):} &   &      \\
MSW (weighted)  & 0.76(0)  & 1.26,0.35,---,---  & 6-56  & 5-56 \\
MSW (unweighted)& .002(0)  & 0.79,0.90,---,---  & 6-57  & 6-57 \\
\multicolumn{3}{l}{\bf All Data:}               &       &      \\
MSW             & 101 (93) & 2.51,0.20,---,---  &       & 4-58 \\
Cosine          & 99.7(91) & 3.16,0.15,1.1,1.1  & 8-12  & 4-58 \\
Sawtooth (6.65) & 97.8(91) & 1.26,0.10,1.8,1.8  & 7-20  & 5-66 \\
Sawtooth (8.05) & 97.4(91) & 1.26,0.05,2.0,2.0  & 5-27  & 5-66 \\
\multicolumn{3}{l}{\bf All Data (averages):}    &       &      \\
MSW (weighted)  & 1.64(0)  & 5.01,0.10,---,---  & 5-20  & 5-55 \\
MSW (unweighted)& 0.15(0)  & 2.51,0.04,---,---  & 6-56  & 6-56 \\
& & & & \\
\end{tabular}
\end{center}
\noindent $^{*}$Parameters: $\Delta m^2/10^{-7} eV^2$,
$\sin^2(2\theta)$,
$A, B (/10^{-10}\mu B$kG), for Zeeman energy = $A + B[\cos(t)$ or
saw$(t)]$.
\caption{Neutrino Data $\chi^2$ Fits and Ga Predictions}
\label{tab:sn-neutrino}
\end{table}

\clearpage

\begin{table}
\begin{center}
\begin{tabular}{lccc}
Model       & $\chi^2$  & Flux reduction  & Ga  \\ \hline
\multicolumn{3}{l}{\bf Concurrent Data:} &     \\
B                &  67.5     & 0.25 of SSM  & 122 \\
B+Be             &  49.4     & 0.30 of SSM  &  98 \\
\multicolumn{3}{l}{\bf Concurrent Data (averages):} & \\
B    (weighted)  &  20.3     & 0.18 of SSM  & 121 \\
B+Be (weighted)  &  9.74     & 0.25 of SSM  &  96 \\
B    (unweighted)&  7.73     & 0.30 of SSM  & 122 \\
B+Be (unweighted)&  2.78     & 0.36 of SSM  & 101 \\
\multicolumn{2}{l}{\bf All Data:}           &     \\
B                &  166      & 0.09 of SSM  & 119 \\
B+Be             &  131      & 0.20 of SSM  &  93 \\
\multicolumn{3}{l}{\bf All Data (averages):}&     \\
B    (weighted)  &  30.4     & 0.07 of SSM  & 119 \\
B+Be (weighted)  &  14.6     & 0.18 of SSM  &  93 \\
B    (unweighted)&  20.0     & 0.15 of SSM  & 120 \\
B+Be (unweighted)&  8.63     & 0.25 of SSM  &  96 \\
\end{tabular}
\end{center}
\caption{Non standard solar model $\chi^2$ fits and Ga predictions}
\label{tab:sn-nssm}
\end{table}

\clearpage

{\bf Figures}

\begin{enumerate}
\item Shown in (a) is the complete Homestake, and Kamiokande
data set used in this analysis, with neutrino signal shown as a fraction
of
that predicted in the Standard Solar Model.  Error bars for
the Cl data are discussed in the text.  In (b) the subset of the sample
containing the data obtained concurrently by the two detectors is shown.
\item Those regions in the MSW parameter space (mass-squared
difference and mixing angle) which are allowed by the 3 year concurrent
data
sample at the 95\% confidence levels based on a comparison to (a) all
the weighted concurrent data, and (b) the unweighted averages of the two
concurrent data sets, are shown. The line shows the solar neutrino
problem
``solution" described by Bahcall and Bethe.
\item Those regions in the MSW parameter space (mass-squared
difference and mixing angle) which are allowed by the full 20 year
weighted
data.
\item  Same as the last figure, except based on (a) the weighted average
signals, (b) the unweighted average signals.
\item  MSW predictions for Homestake and Kamiokande experiments and
experimental rates.
\item Those regions in $\Delta m^2-\sin^22\theta$ space which
are allowed at the 68 and 95\% confidence levels for non-zero transition
magnetic moments based on the 20 year weighted data sample, when the
Zeeman energy is fixed to its ``best fit" value, with time dependence:
\newline
(a) $\left(1.1\times 10^{-10}+1.1\times 10^{-10}\cos(f+kt)\right)\mu
B$kG,
\newline
(b) $\left(2\times 10^{-10}+2\times 10^{-
10}\mbox{saw}(t,\tau=8.05)\right)
\mu B$kG.
\item The predicted electron anti-neutrino signal in Kamiokande
as a fraction of the observed background for incident anti-neutrinos of
energy
$>10.6 MeV$, for resonant spin conversion models, if the Zeeman energy
in the
sun has value: (a) $2\times 10^{-10}\mu B$kG, (b) $5\times 10^{-10}\mu
B$kG,
is shown as a function of $\Delta m^2$ and $\sin^2(2\theta)$.
\end{enumerate}


\clearpage

\begin{thebibliography}{99}
\bibitem{Bahcall1} J. N. Bahcall, ``Neutrino Astrophysics", Cambridge
University
Press (1989);  R. Davis, Jr., in ``Proceedings of the Seventh Workshop
on
Grand Unification", Toyama, 1986, p.237, ed J. Arafune, World
Scientific,
(1986)
\bibitem{Hirata} K. Hirata et al., Phys. Rev. Lett. {\bf 63} (1989) 16;
Phys. Rev. Lett. {\bf 65} (1990) 1297
\bibitem{Filippone} B.W. Filippone and P. Vogel,
Phys. Lett. {\bf B246} (1990) 546
\bibitem{Cleveland} T. Cleveland, Nucl. Instr. Meth. {\bf 214} (1983)
451
\bibitem{Mikheyev} S.P. Mikheyev, A. Yu. Smirnov,
Sov. J. Nucl. Phys. {\bf 42} (1985) 913;
L. Wolfenstein,  Phys. Rev. {\bf D17} (1987) 2369
\bibitem{Bahcall2} J.N. Bahcall and H.A. Bethe,
Phys. Rev. Lett. {\bf 65} (1990) 2233
\bibitem{Voloshin} M.B.Voloshin and M.I. Vysotsky, ITEP Report No. 1
(1986);
L.B. Okun, Sov. J.  Nucl. Phys. {\bf 44} (1986) 546;
L.B. Okun, M.B. Voloshin, and M.I. Vysotsky,
Sov. J.  Nucl. Phys. {\bf 44} (1986) 440;
L.B. Okun, M.B. Voloshin, and M.I. Vysotsky,
Sov. Phys. JETP {\bf 64} (1986) 446;
A. Cisneros, Astrophys. Space Sci. {\bf 10} (1981) 87
\bibitem{Lim1} C.S. Lim and W.J. Marciano, Phys. Rev. {\bf D37} (1988)
1368
\bibitem{Lim2} C.S. Lim et al., Phys. Lett. {\bf B243} (1990) 389
\bibitem{Glashow} see S.L. Glashow and L.M. Krauss, Phys. Lett.
{\bf B190} (1987) 199
\bibitem{Bahcall3} J.N. Bahcall and W. H. Press, Ap. J. {\bf 370} (1991)
730
\bibitem{Krauss1} L.M.Krauss, Nature {\bf 348} (1990) 403
\bibitem{Bieber} J.W. Bieber, D. Seckel, T. Stanev and G. Steigman,
Nature {\bf 348} (1990) 408
\bibitem{Raffelt}  G. Raffelt, Phys. Rev. Lett. {\bf 64} (1990) 2856
\bibitem{Press} W.H. Press, B. P. Flannery, S. A. Teukolsky, W.T.
Vetterling,
``Numerical Recipes", Cambridge University Press, (1986)
\bibitem{Rosen} S.P. Rosen and J.M. Gelb, Phys. Rev. {\bf D34} (1986)
969
\bibitem{Krauss2} L.M. Krauss, Nature {\bf 329} (1987) 689
\bibitem{Nakahata} M. Nakahata, Ph.D Thesis, Search for $^8B$ neutrinos
at
KAMIOKANDE-II, ICEPP Preprint UT-ICEPP-88-1
\bibitem{Barbieri} R. Barbieri et al., Phys. Lett. {\bf B259} (1991) 119
\bibitem{Marciano-review} W. Marciano, Nucl. Phys. {\bf B11} (1989) 5
\bibitem{Krauss-SAGE} L.M. Krauss, Nature {\bf 355} (1992) 399
\bibitem{SAGE} A.I. Abazov et al., Phys. Rev. Lett. {\bf 67} (1991) 3332
\bibitem{pmode} Y. Elsworth et. al., Nature {\bf 347} (1990) 536
\bibitem{cabrera} B. Cabrera, L.M. Krauss, F. Wilczek, Phys. Rev. Lett.
{\bf
55} (1985) 25
\bibitem{wilczek}  L.M. Krauss, F. Wilczek, Phys. Rev. Lett. {\bf 55}
(1985)
122; see also J. Bahcall, IAS preprint 1991

\end{thebibliography}
\end{document}